\documentclass{aastex}
\usepackage{spr-astr-addons}
\usepackage{url}

\RequirePackage{color}

\begin{document}

\title{Quark-Hadron Phase Transition in DGP Brane Gravity with Bulk Scalar Field}
\slugcomment{Not to appear in Nonlearned J., 45.}
\shorttitle{Short article title}
\shortauthors{Autors et al.}

\author{T. Golanbari\altaffilmark{1,a}} \affil{$^{1}$t.golanbari@uok.ac.ir; t.golanbari@.com}
\and
\author{A. Mohammadi\altaffilmark{2,b}} \affil{$^{2}$abolhassanm@gmail.com}
\and
\author{Z. Ossoulian\altaffilmark{3,a}} \affil{$^{3}$zossoulian@gmail.com}
\and
\author{Kh. Saaidi\altaffilmark{4,a}}   \affil{$^{4}$ksaaidi@uok.ac.ir}
\altaffiltext{a}{Department of Physics, Faculty of Science, University of\\ Kurdistan, Pasdaran St., P.O. Box 66177-15175, Sanandaj, Iran.}
\altaffiltext{b}{Young Researchers and elites Club, Sanandaj Branch, \\ Islamic Azad University, Sanandaj, Iran.}

\begin{abstract}
A DGP brane-world framework is picked out to study quark-hadron phase transition problem. The model also includes a bulk scalar field in agreement with string theory prediction. The work is performed using two formalisms as: smooth crossover approach and first order approach, and the results are plotted for both branches of DGP model. General behavior of temperature is the same in these two approaches and it decrease with increasing time and expanding Universe. Phase transition occurs at about micro-second after the big bang. The results show that transition time depends on brane tension value in which larger brane tension comes to earlier transition time.
\end{abstract}



\section{Introduction}
\label{sec:intro}
Recently, brane-world cosmology has attracted a huge attention. The model, which is inspired from string/M theory, was introduced by Randall and Sundrum in 1999 \citep{1,1a}. Based on this picture, the Universe is a four-dimensional hypersurface (the brane) which is embedded in five-dimensional space-time (the bulk). All standard matters and their interactions are confined on the brane, and only gravity could propagate along fifth dimension. This new picture of the Universe brings cosmologists some attractive opportunities to overcome their problems. Field equations and the evolution equations related to this model are derived in \citep{2,2a,2b}. It is shown that the equations contain some correction terms with respect to their corresponding equation in standard four-dimension cosmology. The main correction term is quadratic of energy density which describes a completely different behavior for the Universe in high energy regime. On the other hand, in low energy regime, the quadratic term could be ignored against to the linear term of energy density and comes back to standard form. In 2000, another model of brane-world was proposed by Dvali, Gabadadze and Porrati which is known as DGP brane-world \citep{3}. In this model, the brane is embedded in five-dimension Minkowski space-time with infinite extra dimension. The main difference of DGP model with RS model is the presence of Einstein-Hilbert term in brane action which is known as a quantum correction due to coupling of confined brane matter to bulk gravity \citep{4,4a,5,5a}. Recovery of standard cosmology is somewhat different with respect to RS model. The standard cosmology is recovered in small length scale and for large length scale we encounter modified equation with new behavior for the Universe. This length scale is compared with predicted crossover scale of the model. Cosmological consequences of the model have been studied in \citep{6} which display a self-accelerating solution for the Universe (refer to \citep{5,5a} for other studies in DGP model and \citep{7} for a review).
Additionally, it is well-known that the DGP model is divided to two branches related to the value of $\epsilon$ parameter which appears in the Friedmann equation ($\epsilon=+1$ or $\epsilon=-1$). In brief and for original model of DGP, these two branches can be explained as bellow:
\begin{itemize}
\item \underline{$\epsilon=+1$}: The late time behavior of this case is different. Friedmann equation never goes to five-dimensions regime, and if we assume that the energy density decreases with increasing time, the model describes an inflationary solution. Then, $\epsilon=+1$ branch is able to give a quintessence-like scenario. This branch of the DGP model is known as \textit{self-accelerating branch} \citep{4,4a}.
\item \underline{$\epsilon=-1$}: In late time, under some conditions, the Friedmann equation could describe a fully five-dimensions regime and have a transition from four-dimensions regime to five-dimensions regime. This branch of the DGP model is known as \textit{normal branch} \citep{4,4a}.
\end{itemize}
Between these two branches, the self-accelerating branch undergoes a ghost instability problem. It seems that this problem appears in quantum level.\\

There is another interesting models in standard cosmology, known as scalar field models, which leads to some attractive consequences. The problem of scalar field in brane-world cosmology received huge interest as well. According to string theory suggestion, it is possible to have a scalar field in the bulk which is free to propagate in extra dimension. Therefore, it seems more consistence to have a bulk scalar field instead of the brane scalar field.\\

On the other hand,  according to the  standard model of cosmology,
as the  Universe expanded and cooled it passed through  a series of symmetry-breaking phase transitions which might have  generated  topological defects \citep{8}.   This early Universe  phase transition could have been of first, second, or higher order, and it  has been  studied in detail for over three decades \citep{9,9a,9b,9c,9d,9e,9f,9g}; the possibility of no phase transition was considered  in Ref.\citep{10,10a,10b,10c,10d}.  The phase transition has been considered in standard cosmology by using both first order formalism and crossover formalism \citep{11,11a,11b,11c,11d,11e,11f,11g,11h,11i,11j,11k} with viscosity effect \citep{12,12a,12b,12c,12d} that causes a kind of non-conservation relation. Furthermore, the first order and the crossover approaches were studied in the brane-world \citep{8,13}, DGP brane-world \citep{14},  and in Brans-Dicke models of brane-world gravity \citep{15} without any bulk-brane energy transfer.\\

In the present work, we are going to study the problem of quark-hadron phase transition in early time of the Universe evolution in the DGP brane-world model including bulk scalar field. Difference of Friedmann equation certainly leads to new results about the time of transition, with attention to this fact that presence of bulk scalar field in this model is more consistence with string theory prediction. The problem of phase transition will be considered in two formalisms. In the first formalism, we take smooth crossover approach and study the temperature evolution in high and low temperature regimes. In second case, first order phase transition formalism is taken, and temperature evolution is investigated before, during and after phase transition.
Above topics will be studied for both branches of the DGP model.  Due to ghost instability problem of self-accelerating branch, it might sound that investigation of the phase transition in $\epsilon=+1$ is unnecessary. However, it should be mentioned that this problem appears at the quantum level, where the theory of gravity seems unclear in general as well. Therefore, it is thought that this is so soon to abandon the self-accelerating branch. Also, there are so many other works which they have studied different feature of self-accelerating branch, which it means that this branch still has scientists  interests. First, phase transition will be considered for self-accelerating branch. It is trying to clearly explain every step of the work. The main equations and processes of the work, are expressed generally which could be used for both branches and only different is in the value of $\epsilon$. In second stage, phase transition will be studied for normal branch. Since the general equations have already derived, it is preferred to ignored the detail, and {the results are discussed}. \\

This paper is organized as follows: In Sec. 2, we introduce the model and derive the main equations.
We study quark-hadron phase transition in high and low temperature regimes in our model using smooth crossover approach; and we review the first-order phase transition and consider it in our model toward investigate of temperature evolution before, during and after phase transition for $\epsilon=1$ in Sec. 3, and for $\epsilon=-1$ in Sec. 4 and Sec. 5 summarizes our results and compares with the previous results of other works.

\section{Field Equation of Brane}
 In such a model, the action commonly contains two main parts as bulk and brane action
\begin{equation}\label{II.01}
S =  S_{\rm bulk}+ S_{\rm brane}.
\end{equation}
The bulk action is defined as
\begin{equation}\label{II.02}
S_{\rm bulk} =\int_{\cal M} d^5X \sqrt{-{}^{(5)}g} \left[ {1 \over 2 \kappa_5^2} {}^{(5)}R + {}^{(5)}L_{\rm B} \right],
\end{equation}
 where $^{(5)}L_{\rm B}$ indicates the matter of bulk. It is assumed that the bulk is filled by two components, as cosmological constant $\Lambda$, and scalar field $\phi$; given by $^{(5)}L_{\rm B}=-1 / 2 nabla_A\phi\nabla^A\phi - V(\phi) - \Lambda$.  ${}^{(5)}R$ is five-dimensional Ricci scalar related to five-dimensional metric ${}^{(5)}g_{AB}$. $\kappa_5^2$ is related to five-dimensional Planck mass by $\kappa_5^2=8\pi G_5=M_5^{-3}$, where $ G_5 $ is the five-dimensional Newtonian gravitational constant. The bulk is filled by two component as cosmological constant $\Lambda$ and scalar field $\phi$.  \\
 Additionally, the action of the brane is described as following
\begin{equation}\label{II.03}
S_{\rm brane}=\int_{M} d^4 x\sqrt{-g}\left[ [\tilde{K}] + L_{\rm brane}(g_{\alpha\beta},\psi) \right],
\end{equation}
where $ \tilde{K} = {[K] / \sqrt{-g} \kappa_5^2} $; and $ [K] $ is exterior curvature on each side of brane and is known as Gibbons-Hawking term. $L_{\rm brane}$ is general form of brane matter Lagrangian. $g_{\mu\nu}$ is induced metric on brane defining by $g_{\mu\nu}=\delta^A_{\ \mu}\delta^B_{\ \nu}\; ^{(5)}g_{AB}$. Taking variation of action with respect to five-dimensional metric leads one to the five-dimension Einstein field equations
\begin{equation}\label{II.04}
{}^{(5)}G_{AB} =  \left[ \,{}^{(5)}T_{AB} +\tau_{AB}\, \delta(y)\,\right].
\end{equation}
${}^{(5)}T_{AB}$ indicates the total bulk energy momentum tensor and are defined as
\begin{equation}\label{II.05}
{}^{(5)}T_{AB} \equiv -2 {\delta {}^{(5)}\!L_{\rm B} \over \delta {}^{(5)}g^{AB}}  +{}^{(5)}g_{AB}{}^{(5)}\!L_{\rm B}= {}^{(5)}T^{(\phi)}_{AB} +{}^{(5)}T^{(\Lambda)}_{AB}.
\end{equation}
Note that in above relations we set $\kappa_5=1$, and we  use this convention for the rest of this work.  Each component of the bulk energy-momentum tensor is derived as
\begin{equation}\label{II.07}
{}^{(5)}T^{(\phi)}_{AB}= \nabla_A\phi \nabla_B\phi - g_{AB} \big( \frac{1}{2}g^{CD} \nabla_C\phi \nabla_D\phi + V(\phi) \big),
\end{equation}
\begin{equation}\label{II.08}
{}^{(5)}T^{(\Lambda)}_{AB}=-\Lambda g_{AB}.
\end{equation}
{ The effective energy-momentum tensor of brane in the field equations (\ref{II.04}) is denoted by $\tau_{\mu\nu}$, which is illustrated as }
\begin{equation}\label{II.09}
\tau_{\mu\nu}\equiv -2 {\delta L_{\rm brane} \over \delta g^{\mu\nu}}  +g_{\mu\nu}L_{\rm brane}.
\end{equation}

Following \citep{16}, the effective field equations of the brane are resulted as 
\begin{eqnarray}
G_{\mu\nu} = {2  \over 3}&\biggl[& g_{\mu\nu} \biggl({}^{(5)}T_{RS}~n^R n^S -{1 \over4}{}^{(5)}T\biggr)\nonumber \\
 & + & {}^{(5)}T_{RS}~g^{R}_{~\mu}  g^{S}_{~\nu} \biggr] + \kappa_5^4\pi_{\mu\nu} - E_{\mu\nu}, \label{II.10}
\end{eqnarray}
where $n^A$ is unit normal vector, and $\pi_{\mu\nu}$  is expressed by
\begin{equation}\label{II.11}
\pi_{\mu\nu}= -\frac{1}{4} \tau_{\mu\alpha}\tau_\nu^{~\alpha} + \frac{1}{12}\tau\tau_{\mu\nu} + \frac{1}{8} g_{\mu\nu} \tau_{\alpha\beta} \tau^{\alpha\beta} - \frac{1}{24} g_{\mu\nu}\tau^2\,
\end{equation}
 Above relation contains quadratic term of energy-momentum tensor.  Consequently, it produces quadratic term of energy density, which is the main difference of brane world scenario with standard four dimension model. $E_{\mu\nu}$ is another modified term of the field equations Eq.\;(\ref{II.10}), which is picture of bulk Weyl tensor on brane including some informations about bulk gravitation 
\begin{equation}\label{II.12}
E_{\mu\nu} = {}^{(5)}C_{MRNS}~n^M n^N g_{~\mu}^{R}~ g_{~\nu}^{S}.
\end{equation}

 Scalar field equation of motion is another dynamical equation of the model obtained by taking variation of the action with respect to scalar field
\begin{equation}\label{II.13}
{}^{(5)}\Box \phi = V'(\phi),
\end{equation}
where ${}^{(5)}\Box$ indicates five-dimensional D'Alambert, and prime denotes derivative with respect to scalar field.
\subsection{DGP brane world scenario}

 The interaction between matter on the brane and the bulk gravity produces some quantum correction on the brane, resulted in induced gravity on brane. Based on Dvali, Gabadadze and Porrati, this quantum correction appears as Ricci scalar in brane action\citep{3}
\begin{equation}\label{II.14}
L_{\rm brane}=  {1 \over 2\mu^2} R -  \lambda + L_{\rm m}.
\end{equation}
where $ \mu^2=M_4^{-2} $, and $ M_4 $ is four-dimensional Planck mass. Note  that, it is a generalized version of original DGP model; obtained by taking $\lambda=\Lambda=0$.  Then, the effective brane energy-momentum tensor is derived
\begin{equation}\label{II.15}
\tau^{\mu}_{~\nu}=-\lambda \delta^{\mu}_{~\nu} +T^{\mu}_{~\nu}-\mu^{-2} G^{\mu}_{~\nu}.
\end{equation}
Using Eq.\;(\ref{II.15}), one can find out the effective field  equations of brane
\begin{eqnarray}\label{II.16}
\left(1+{\lambda \over 6\mu^2}\right)
G_{\mu\nu} & = & -\Lambda_4 g_{\mu\nu} + \left[\pi_{\mu\nu}^{(T)} +\mu^{-4} \pi^{(G)}_{\mu\nu}\right] \\
 & + & {\lambda\over 6} T_{\mu\nu} + \mu^2 \tilde{T}_{\mu\nu} + {1 \over \mu^2} \mathcal{G}_{\mu\nu}-E_{\mu\nu}, \nonumber
\end{eqnarray}
where
\begin{eqnarray}\label{II.17}
\Lambda_4 &=& {1 \over 2} \Big[  \Lambda + {1 \over 6}  \lambda^2\Big]
\end{eqnarray}
\begin{eqnarray}\label{II.18}
\pi_{\mu\nu}^{(T)} = + \frac{1}{4} T_{\mu\alpha}T_\nu^{~\alpha}+\frac{1}{12}TT_{\mu\nu}& + &\frac{1}{8}g_{\mu\nu}T_{\alpha\beta}T^{\alpha\beta} \nonumber \\
& - & \frac{1}{24}g_{\mu\nu}T^2,
\end{eqnarray}
\begin{eqnarray}\label{II.19}
\pi_{\mu\nu}^{(G)} = -\frac{1}{4} G_{\mu\alpha}G_\nu^{~\alpha}+\frac{1}{12}GG_{\mu\nu} & + & \frac{1}{8}g_{\mu\nu}G_{\alpha\beta}G^{\alpha\beta}\nonumber \\
 & - & \frac{1}{24}g_{\mu\nu}G^2
\end{eqnarray}
\begin{eqnarray}\label{II.20}
 \mathcal{G}_{\mu\nu} = & & {1 \over 4} \big( G_{\mu\rho}\tau_\nu^{\ \ \rho} + \tau_{\mu\rho}G_\nu^{\ \ \rho} \big) - {1 \over 12} \big( \tau G_{\mu\nu} + G\tau_{\mu\nu} \big)\nonumber \\
  & - & {q_{\mu\nu} \over 2} \big( G_{\alpha\beta}\tau^{\alpha\beta} - {1 \over 3} G\tau \big)
\end{eqnarray}

$\Lambda_4$ is the effective cosmological constant of  the brane, which same as RS-II model could be taken as zero. $\tilde{T}_{\mu\nu}$ is  the bulk scalar field energy-momentum tensor, which depicts the effect of the bulk on the brane evolution. This tensor is expressed by
\begin{eqnarray}\label{II.21}
\tilde{T}_{\mu\nu} = {l_{DGP} \over 3}\biggl( 4\phi_{,\mu}\phi_{,\nu}+ \Big[ {3 \over 2}(\phi_{,5})^2
& - & {5 \over 2}g^{\alpha\beta}\phi_{,\alpha}\phi_{,\beta}
\nonumber \\
& - & 3V(\phi) \Big]g_{\mu\nu}\biggl),
\end{eqnarray}
where $l_{DGP}=\kappa_5^2 / 2\mu^2$ is the crossover length scale between the 4D and 5D regimes in the DGP brane model. The conservation equation is  given by
\begin{equation}\label{II.22}
D^\nu T_{\mu\nu}=0 \qquad \Longrightarrow \qquad \dot{\rho}+3H(\rho+p)=0,
\end{equation}
where $D$ indicates covariant derivative with respect to $g_{\mu\nu}$ metric.
\subsection{Evolution Equation}

 The (0-0)-component of field equations is
\begin{eqnarray}\label{II.23}
\left(1+{\lambda \over 6\mu^2}\right) G_{00} =  -\Lambda_4 g_{00} & + & \left[\pi_{00}^{(T)} +\mu^{-4} \pi^{(G)}_{00}\right]\nonumber \\
& + & \mu^2 \tilde{T}_{00} + {\lambda\over 6} T_{00}\nonumber \\
& + & { \mathcal{G}_{00}\over \mu^2}-E_{00}.
\end{eqnarray}

Following \citep{1}, and for simplicity, the four-dimensional cosmological constant is taken zero. The spatially flat FLRW metric is picked out in the rest of the work
\begin{equation}\label{II.24}
ds^2 = -dt^2 + a^2(t) \delta_{ij}dx^i dx^j + dy^2.
\end{equation}
Substituting the metric in Eqs.\;(\ref{II.18}), (\ref{II.19}), (\ref{II.20}) and (\ref{II.21}) leads one to
\begin{eqnarray}
\tilde{T}_{00} &=& \rho_B = l_{DGP} \big( \dot\phi^2 /2 + V(\phi) \big) \label{II.25}\\
\pi_{00}^{(T)}&= & {\rho^2 \over 12} \label{II.26}\\
\pi_{00}^{(G)}&= & {1 \over 12} G_{00}G^{00} \label{II.27}\\
 \mathcal{G}_{00} & =& -{1 \over 6} \rho G_{00} \label{II.28}
\end{eqnarray}
where $G_{00}=3H^2$. To derive Eq.\;(\ref{II.25}), we impose a boundary condition as $\partial_y \phi |_{y=0}=0$ \citep{17,17a,17b}.  Utilizing the above expressions, the dynamical equation of the model is read from Eq.\;(\ref{II.23})
\begin{eqnarray}\label{II.29}
 - {3 \over 4\mu^4} H^4 & + & 3 \left(1+{1 \over 6\mu^2}(\rho + \lambda) \right)H^2\nonumber \\ & - &\mu^2 \rho_B - {\lambda \over 6}\rho - {1 \over 12} \rho^2 + {\zeta \over a^4} = 0.
\end{eqnarray}
Rearranging Eq.(\ref{II.29}), the modified Friedmann equation is derived
\begin{eqnarray} \label{II.30}
{H^2 \over 2\mu^4} =  \left[ \chi + \epsilon \sqrt{\chi^2 - {1 \over 3\mu^4} \Bigg[ \mu^2 \rho_B + {\lambda \over 6}\; \rho \left( 1 + {\rho \over 2\lambda} \right) \Bigg] } \ \ \right]
\end{eqnarray}
where $\chi = ( 1 + (\rho+\lambda) / 6\mu^2 )$, and $\epsilon = \pm 1$. It should be noted that the contribution of dark radiation has been ignored finding above relation.

To consider phase transition, we need to know the function of scalar field. The exact form of scalar field is derived by solving Eq.\;(\ref{II.13}); although finding the solution encounter with difficulty. However, one can apply some
appropriate  assumptions to derive scalar field. Based on inflationary scenario, scalar field potential dominates the universe and a quasi-de Sitter expansion  occurs, while scalar field  moves very slowly to the minimum of its potential. After inflation, scalar field begin to oscillate in minimum of potential, and other standard particle could be produced. Therefore, one can conclude that, the potential of scalar field could be ignored after inflation and reheating  era of the Universe evolution. According to this argument, the five-dimensional scalar field equation of motion on brane could be read as
\begin{equation} \label{II.31}
\ddot\phi + 3H\dot\phi - \partial_{yy}\phi + V'(\phi) = 0.
\end{equation}
To solve above differential equation, one need to assume an ansantz for $\partial_{yy}\phi$. According to \citep{18}, we take it as $\partial_{yy}\phi = -\ddot\phi$.  Then, the scalar field is derived as $\dot\phi^2 /2 = \phi_0 a^{-3}$, where $\phi_0$ is constant of  integration; in another word, the bulk energy density on brane can be expressed by
\begin{equation}\label{II.32}
\rho_B = l_{DGP} \phi_0 a^{-3}.
\end{equation}

\section{Quark-Hadron phase transition in $\epsilon=+1$ Branch of the DGP model}
As beginning, we are going to study the $\epsilon=+1$ branch of the model for quark-hadron phase transition.\\


\subsection{Lattice QCD Phase Transition}

One of definite prediction of QCD is the existence of a phase transition from a quark-gluon plasma phase to hadron gas phase. The transition could be first, second or higher order; or it only can be a crossover with a rapidly change in some of observable which strongly depend on the values of quark masses. Todays, there is two kind of phase transition which is popular among scientist and seen in papers:1. First order phase transition, 2. Crossover transition. In this section, we are going to study phase transition utilizing smooth crossover approach.

One of  the fundamental concepts in particle physics is QCD phase transition, and incredibly becomes relevant to any study which concern early universe. Recently lattice QCD calculation for two quark flavors suggest that QCD makes a smooth crossover.

Lattice QCD is a new approach which allows one to systematically study the non-perturbative regime of the QCD equation of state.  By using supercomputers, the QCD equation of state was computed on the  lattice \citep{19,19a} with two light quarks and a heavier strange quark. In \citep{19,19a}, the data for energy density $\rho(T)$, pressure $p(T)$, trace anomaly $\rho-3p$ and entropy $s$ have been formed;  and in the current work, we only use this data. Recent information on lattice QCD in high temperature could be found in \citep{21,21a,21b,21c}, which emphasize that in high temperature, as it was expected, there is radiation like behavior (equations (\ref{III.02}) and (\ref{III.03}) in the main paper).

The trace anomaly can be accurately calculated in the high temperature region, while in the low temperature region, it is affected by possibly large discretization effects. Therefore to construct realistic equation of state, we could use the lattice data for the trace anomaly in the high temperature region, $T \geq 250$MeV, and use  Hadronic Resonance Gas (HRG) model in the low temperature region, $T \leq 180$MeV. (for more detail refer to \citep{22,22a}). The HRG result for trace anomaly can be performed by the simple form as \citep{21c}
\begin{equation}\label{III.01}
{I(T) \over T^4} \equiv {\rho-3p \over T^4} = a_1 T + a_2 T^3 + a_3 T^4 + a_4 T^{10}.
\end{equation}

In lattice QCD, the calculation of the energy density, pressure, and entropy density usually proceed through the calculation of the trace anomaly. Finally the energy density and pressure could be obtained as
\begin{equation}\label{III.02}
\rho(T)= 3a_0T^4 + 4a_1T^5 + 2a_2T^7 + {7a_3 \over 4}T^8 + {13a_4 \over 10} T^{14},
\end{equation}
\begin{equation}\label{III.03}
p(T) =a_0T^4 + a_1T^5 + {a_2 \over 3}T^7 + {a_3 \over 4}T^8 + {a_4 \over 10} T^{14}.
\end{equation}

In the following subsections, we are going to study phase transition using crossover approach for two main regime as high temperature regime and low temperature regime.\\
\subsubsection{High Temperature Regime}

As it was mentioned above, in high temperature regime one can use lattice QCD data for trace anomaly in order  to find realistic equation of state. It is found out that there is a radiation like behavior and we have a simple form of equation of state as
\begin{eqnarray}
\rho(T) & \simeq & \alpha T^4 , \label{III.04}\\
p(T) & \simeq & \sigma T^4 , \label{III.05}
\end{eqnarray}
where $\alpha = 14.9702\pm 009997$ and $\sigma = 4.99115\pm 004474$. Inserting Eqs.\;(\ref{III.04}) and (\ref{III.05}) in conservation equation (\ref{II.22}), the Hubble parameter and scale factor could be obtained as
\begin{equation}\label{III.06}
H=- {4\alpha \over 3(\alpha+\sigma)} {\dot{T} \over T} \qquad , \qquad a(T)=cT^{-{4\alpha \over 3(\alpha+\sigma)}}.
\end{equation}

\begin{figure}[h]
\centerline{\includegraphics[width=7cm] {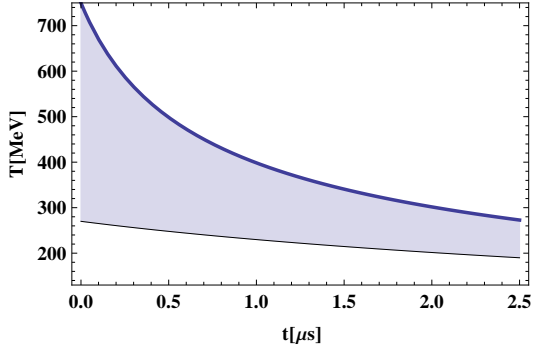}}
\caption{\label{HT} {\small $T$ versus $t$ according to high temperature region
 (for $ 250<T<750 \ {\rm MeV} $) of the smooth crossover procedure in the DGP brane
 gravity including bulk scalar field. The constant parameters are taken as:
$\lambda = 10^{10} {\rm MeV^4}$, $l_{DGP} = 10^{11} {\rm MeV^{-1}}$,
$\kappa_5 = \sqrt{2\mu^2 l_{DGP}}=1 {\rm MeV^{-3/2}}$,
$\phi_0 = 2 \times 10^5 $ and $\epsilon = 1$.  }}
\end{figure}

Substituting above relations in Friedmann equation (\ref{II.30}), leads to the differential equation of  temperature
\begin{eqnarray}\label{III.07}
\dot{T}& = - &{3(\alpha+\sigma)T \over 4\alpha} \times \biggl\{ 2\mu^4  \Bigg[ \chi \\
& & + \epsilon \sqrt{ \chi^2 - {1 \over 3\mu^4} \Big[ \mu^2 \rho_B + {\lambda \over 6}\; \rho \left( 1 + {\rho \over 2\lambda} \right) \Big] } \Bigg] \biggl\}^{\frac{1}{2}},\nonumber
\end{eqnarray}
where displays the behavior of temperature with respect to the cosmic time. The differential equation (\ref{III.07}) is solved numerically and the results have been depicted in Fig.(\ref{HT}). It could be found out that the phase transition occur at about $2-2.5$ micro-second after big bang. Transition time depends on the brane tension value in which larger transition time is obtained for larger values of brane tension.\\

\subsubsection{Low Temperature Regime}

\begin{figure}[h]
\centerline{ \includegraphics[width=7cm] {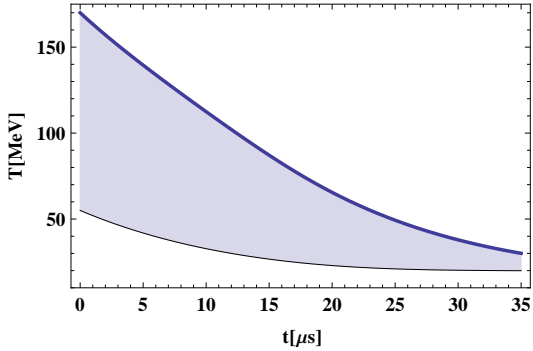}}
\caption{\label{LT} {\small $T$ versus $t$ according to low temperature region
(for $ 50<T<180 \ {\rm MeV}$) of the smooth crossover procedure in the DGP brane gravity
including bulk scalar field. The constant parameters are taken as:
$\lambda = 10^{10} {\rm MeV^4}$, $l_{DGP} = 10^{11} {\rm MeV^{-1}}$,
$\kappa_5 = \sqrt{2\mu^2 l_{DGP}}=1 {\rm MeV^{-3/2}}$,
$\phi_0 = 2 \times 10^5$ and $\epsilon = 1$. }}
\end{figure}

In the low temperature regime, the situation is different with high temperature regime, so that the trace anomaly is not a suitable approach to obtain equation of state because of the fact that it is affected by large discretization effects. However the HRG is still a good model  to achieve a realistic equation of state as mentioned.
In the HRG scenario, the confinement phase of QCD  is treated
as a non-interacting gas of fermions and bosons \citep{22}.  The idea of the HRG model is to
implicitly account for the strong interaction in the confinement phase by looking at the hadronic
resonances only, since these are basically the only relevant degrees of freedom in that phase. The results of HRG is  parameterized for trace anomaly so that
\begin{equation}\label{III.08}
{I(T) \over T^4}\equiv  {\rho -3p \over T^4} = a_1T + a_2T^3 + a_3T^4 + a_4T^{10},
\end{equation}
where $a_1$ = 4.654 GeV$^{-1}$, $a_2$ = -879 GeV$^{-3}$, $a_3$ = 8081 GeV$^{-4}$, $a_4$ = -7039000 GeV$^{-10}$. Through the calculation of the trace anomaly $I(T)=\rho(T) - 3p(T)$ in lattice QCD, the parameters energy density, pressure, entropy density is estimated using the usual thermodynamics identities. The difference of pressure at two temperature is expressed as the integral of trace anomaly which is described by following relation
\begin{equation}\label{III.09}
{p(T) \over T^4 } - {p(T_{\rm low})\over T^4_{low}} = \int^T_{T_{\rm low}} {dT' \over T'^5}I(T') ,
\end{equation}
where $p(T_{\rm low})$ could be ignored because of exponential suppression for sufficiently small value of lower integration limit. On the other hand, the energy density could be obtained as $\rho(T) = I(T) + 3p(T)$. Therefore, the energy density and pressure in this regime is expressed by
\begin{equation}
\rho(T) = 3\eta T^4 + 4a_1 T^5 + 2a_2 T^7 + {7a_3 \over 4}T^8 + {13a_4 \over 10} T^{14} , \label{III.10}
\end{equation}
\begin{equation}
p(T) =   \eta T^4 + a_1 T^5 + {a_2 \over 3} T^7 + {a_3 \over 4}T^8 + {a_4 \over 10} T^{14} , \label{III.11}
\end{equation}
where $a_0= -0.112$. Now, let's consider the behavior of temperature. In this case we are actually considering the times before phase transition where the Universe is in confinement phase and is treated as a non-interacting gas of fermions and bosons. Using conservation equation, the Hubble parameter and scale factor can be derived respectively as
\begin{equation}\label{III.12}
H  =  - {12\eta T^3 + 20a_1T^4 + A(T) \over 3\big[ 4\eta T^4 + 5a_1T^3 + B(T) \big]} \dot{T} ,
\end{equation}
\begin{equation}\label{III.13}
a(T) = {c T^{-1} \over\big[ 60 \eta + 75a_1T + 35a_2 T^3 + 30a_3 T^4 + 21 T^10 \big]^{1/3}} ,
\end{equation}
where $A(T)=14 a_2 T^6 + 14 a_3 T^7 + {91 \over 5} T^{13}$ and $B(T)={7 \over 3} a_2 T^7 + 2 a_3 T^8 + {7 \over 5} T^{14}$.  Then from the Friedmann equation (\ref{II.30}), the differential equation of temperature could be derived as
\begin{eqnarray}\label{III.14}
\dot{T} & = - & {3\big[ 4\eta T^4 + 5a_1T^3 + B(T) \big] \over 12\eta T^3 + 20a_1T^4 + A(T)} \times \biggl\{ 2\mu^4 \Bigg[ \chi  \\
 & & \pm \sqrt{ \chi^2 - {1 \over 3\mu^4}
  \Big[ \mu^2 \rho_B + {\lambda \over 6}\; \rho \left( 1 + {\rho \over 2\lambda} \right) \Big] } \ \Bigg] \biggl\}^{\frac{1}{2}} , \nonumber
\end{eqnarray}
which expresses the behavior of temperature as a function of cosmic time. The numerical results of the differential equation (\ref{III.14}) is depicted in Fig.(\ref{LT})  indicating that phase transition occurs at about $30-35$ micro-second after big bang. This transition time occurs after transition time related to high temperature regime which exhibits a consistence result. Transition time in low temperature regime strongly depends on brane tension value so that it increases by enhancement of the brane tension value.
\subsection{First Order Phase Transition}
The HRG model in the temperature interval $180 {\rm MeV} < T < 250 {\rm MeV}$ is no longer valid, then the study of quark-hadron phase transition in this temperature interval should be performed using another formalism. Quark-hadron phase transition in QCD is characterized by the singular behavior of partition function, and might be first or second order phase transition \citep{23}. In this section, we are going to pick out first order phase transition formalism and study the behavior of temperature in DGP brane gravity framework with bulk scalar field in this temperature interval. Based on \citep{23}, the equation of state for matter in quark-gluon phase is given by
\begin{equation}\label{III.15}
\rho _q=3a_q T^4 +U(T), \ \ \ \ \ p_q =a_q T^4 -U(T) ,
\end{equation}
where the subscript $q$ indicates the quark-gluon phase of matter, and the constant $a_q$ is given as $a_q=61.75(\pi^2/90)$. $U(T)$ denotes potential energy density and is expressed by \citep{23}
\begin{equation}\label{III.16}
U(T)=B+\gamma_T T^2 - \alpha_T T^4 ,
\end{equation}
where $B$ is the bag pressure constant, $\alpha_T = 7\pi^2/20$, and $\gamma_T = m^2_s/4$. $m_s$ indicates the mass of the strange quark, which  is in the
$60 - 200$ MeV range. This form of $U$ comes from a model in which the quark fields interact with a chiral field formed by  the $\pi$ meson field together with a scalar field. Results obtained from low energy hadron spectroscopy, heavy ion collisions, and  from phenomenological fits of light hadron
properties give a value of $B^{1/4}$ between 100 and 200 MeV \citep{8}.

After quark-gluon phase, the matter comes to hadron phase, then the cosmological fluid can be taken as an ideal gas of massless pions and nucleons described by the Maxwell-Boltzmann  distribution function with energy density $\rho_h$ and pressure $p_h$. The equation of state for matter in this area of the Universe evolution is expressed by following simple relations
\begin{equation}\label{III.17}
 p_h ={1 \over 3}\rho _h=a_{\pi} T^4 ,
\end{equation}
where $a_{\pi} =17.25 (\pi^2/90)$. The critical temperature $T_c$ is defined by the condition $p_q (T_c) = p_h (T_c)$ \citep{24}. Taking   $m_s = B^{1/4} = 200$ MeV, the critical temperature  is
\begin{equation}\label{III.18}
 T_c={\Bigg[{\frac{\gamma_T + \sqrt{\gamma^2_T+ 4B(a_q+\alpha_T-a_{\pi})}}{2(a_q+\alpha_T -a_{\pi})}}\Bigg]}^{1\over 2}\approx 125 \ {\rm MeV}.
\end{equation}
Since the phase transition is  first order, all physical quantities, such as  the energy density, pressure, and entropy, exhibit discontinuities across the critical curve.

\subsubsection{Behavior of Temperature}

\begin{figure}[h]
\centerline{ \includegraphics[width=7cm] {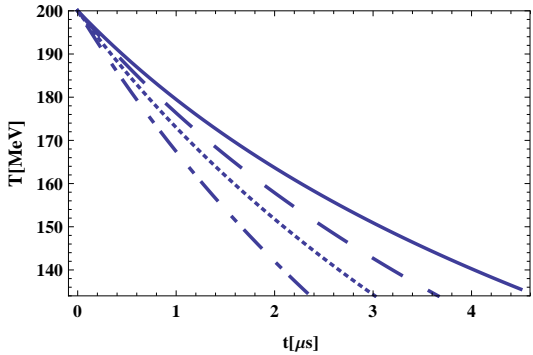}}
\caption{\label{F01} {\small   $T$ versus $t$ in the quark-gluon phase with general
form of potential energy density has been plotted for different values brane tension
$\lambda$ as: $1 \times 10^9$ (solid line), $5 \times 10^9$ (dashed line),
$10 \times 10^9$ (dotted line), $20 \times 10^9$ (dotted-dashed line). The other
constant parameters are taken as:
$l_{DGP} = 10^{11} {\rm MeV^{-1}}$,
$\kappa_5 = \sqrt{2\mu^2 l_{DGP}}=1 {\rm MeV^{-3/2}}$,
$\phi_0 = 2 \times 10^5$, $\epsilon = 1$ and $B^{1/4}=200 \ {\rm MeV}$.}}
\end{figure}

At first stage of phase transition, we study the behavior of temperature when the matter is in quark-gluon phase.  In this regards, we need a specific form of equation of state. Substituting Eqs.\;(\ref{III.15}) and (\ref{III.16}) in conservation equation (\ref{II.22}), one can obtain the Hubble parameter as a function of temperature and its first time derivative
\begin{equation}\label{III.19}
H = - \Big[ {3a_q-\alpha_T \over 3a_q} + {\gamma_T \over 6a_qT^2}\Big] {\dot{T} \over T},
\end{equation}
where  the general form of potential $U(T)$ has been taken. Then, integrating above relation gives scale factor as
\begin{equation}\label{III.20}
a(T) = c T^{{\alpha_T - 3a_q \over 3a_q}} \exp\Big( {\gamma_T \over 12a_q T^2} \Big).
\end{equation}
Therefore, one can easily derive  the temperature evolution equation from Eq.\;(\ref{III.19}) in Friedmann equation (\ref{II.30})
\begin{eqnarray}\label{III.21}
\dot{T} & = & -{6a_q T^3 \over \left[ 2(3a_q-\alpha_T)T^2 + \gamma_T \right]}\times \Bigg\{ 2\mu^4  \Bigg[ \chi \\
 & & \pm \sqrt{ \chi^2 - {1 \over 3\mu^4}\Big[ \mu^2 \rho_B + {\lambda \over 6}\; \rho \left( 1 + {\rho \over 2\lambda} \right) \Big] } \ \Bigg] \Bigg\}^{\frac{1}{2}} \nonumber
\end{eqnarray}
Above expression governs on the evolution of temperature with respect to cosmic time. The differential equation is solved numerically and the results are plotted in Fig(\ref{F01}). It could be realized that the quark-gluon phase is finished at about $2.5-4.5$ micro-second after the big bang which is in agreement with the results of previous section. Earlier transition time occurs for higher values of brane tension value, which expresses dependence of transition time on brane tension.\\

\begin{figure}[h]
\centerline{ \includegraphics[width=7cm] {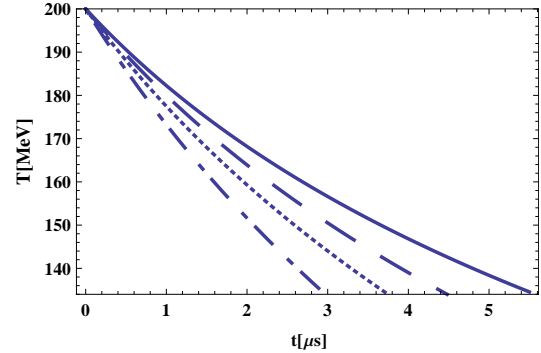}}
\caption{\label{F02} {\small   $T$ versus $t$ in the quark-gluon phase with constant
potential energy density has been plotted for different values brane tension $\lambda$
as: $1 \times 10^9$ (solid line), $5 \times 10^9$ (dashed line), $10 \times 10^9$
(dotted line), $20 \times 10^9$ (dotted-dashed line). The other constant parameters
are taken as: $l_{DGP} = 10^{11} {\rm MeV^{-1}}$,
$\kappa_5 = \sqrt{2\mu^2 l_{DGP}}=1 {\rm MeV^{-3/2}}$,
$\phi_0 = 2 \times 10^5$, $\epsilon = 1$ and $U(T)=B=200 \ {\rm MeV^4}$. }}
\end{figure}

\noindent \textbf{Bag Model}\\
Elastic bag model, which allows quark to move around freely, is one of popular models dealing with quark confinement. In this case, the potential energy density is a constant, namely $U(T)=B$, and equation of state of quark matter describes
by $p_q = (\rho_q - 4B)/3$.  Keeping the assumption that bag model provides the quark equation of state, the Hubble parameter is derived as
\begin{equation}\label{III.22}
H = - {\dot{T} \over T} \qquad \Longrightarrow \qquad a(T) = {c \over T}.
\end{equation}
Utilizing the Friedmann equation (\ref{II.30}) leads to following relation
\begin{eqnarray}\label{III.23}
T = & -T & \biggl\{2\mu^4  \Bigg[ \chi \\
& + & \epsilon \sqrt{ \chi^2 - {1 \over 3\mu^4} \bigg[ \mu^2 \rho_B + {\lambda \over 6}\; \rho \Big( 1 + {\rho \over 2\lambda} \Big) \bigg] }\Bigg] \biggl\}^{\frac{1}{2}},\nonumber
\end{eqnarray}
which describes time evolution of temperature as a function of cosmic time. The numerical results of the differential equation has been plotted in Fig.(\ref{F02}). { In comparison}, phase transition in bag model occur later than general case and quark-gluon phase is finished at about $3-5.5$ micro-second after the big bang. Dependence on brane tension is clear from Fig.(\ref{F02}) which show earlier transition for higher values of brane tension.

\subsubsection{Evolution of hadron Volume Fraction}

During phase transition, the matter energy density decreases from quark energy density, $\rho_Q$, to hadron energy density, $\rho_H$. In this time, pressure, enthalpy and entropy remain conserved, and temperature at $T=T_c= 125 {\rm MeV}$ is conserved as well.  For the constant parameters, there are $\rho_Q = 5 \times 10^9 $ MeV$^4$, $\rho_H \approx 1.38 \times 10^9$ MeV $^4$, and constant pressure $p_c \approx 4.6 \times 10^8 $ MeV$^4$. Following \citep{8,14,15}, the parameter $\rho(t)$ could be replaced by volume fraction of matter in hadron phase, $h(t)$, defined by
\begin{equation}\label{III.24}
\rho = \rho_H h(t) + \big( 1-h(t) \big)\rho_Q
\end{equation}
where $\rho_H$ and $\rho_Q$ respectively are energy density of hadron and quark. The Hubble parameter could be derived from conservation equation
\begin{equation}\label{III.25}
H=-{1 \over 3} {(\rho_H - \rho_Q)\dot{h} \over \rho_Q+p_c+(\rho_H-\rho_Q)h}
= -{1 \over 3} {r\dot{h} \over 1+rh}
\end{equation}
where $r=(\rho_H - \rho_Q) / (\rho_Q+p_c)$.  Integrating Eq.\;(\ref{III.25}) gives the scale factor as a function of the hadron volume fraction
\begin{equation}\label{III.26}
a(t) = a(t_c) \big[ 1+rh(t) \big]^{-1/3}
\end{equation}
where we  assumed $ h(t_c)=0 $. So, plugging Eq.\;(\ref{III.25}) into Friedmann equation (\ref{II.30}), the time evolution equation of the matter fraction in the hadronic phase is
\begin{eqnarray}\label{III.27}
\dot{h} = & - & 3 {(1+rh) \over r} \times \biggl \{ 2\mu^2  \Bigg[ \chi \\
& + & \epsilon \sqrt{ \chi^2 - {1 \over 3\mu^4}
 \Big[ \mu^2 \rho_B + {\lambda \over 6}\; \rho \left( 1 + {\rho \over 2\lambda} \right) \Big] } \ \ \Bigg] \biggl \}^{\frac{1}{2}}.\nonumber
\end{eqnarray}
\begin{figure}[h]
\centerline{ \includegraphics[width=7cm] {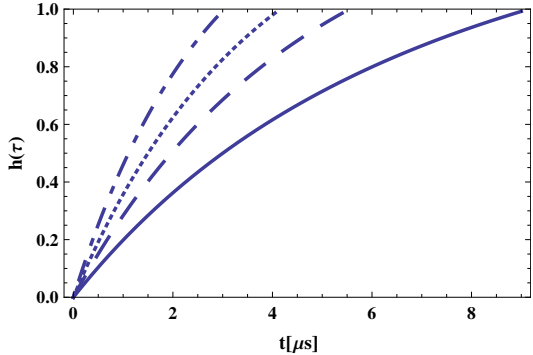}}
\caption{\label{F03} {\small   Variation of hadron fraction parameter $h$ versus $t$
has been depicted for different values brane tension $\lambda$ as:
$1 \times 10^9$ (solid line), $5 \times 10^9$ (dashed line), $10 \times 10^9$
(dotted line), $20 \times 10^9$ (dotted-dashed line). The other constant parameters
are taken as: $\lambda = 10^{10} {\rm MeV^4}$, $l_{DGP} = 10^{11} {\rm MeV^{-1}}$,
$\kappa_5 = \sqrt{2\mu^2 l_{DGP}}=1 {\rm MeV^{-3/2}}$,
$\phi_0 = 2 \times 10^5 $ and $\epsilon = 1$. }}
\end{figure}
\begin{figure}[h]
\centerline{ \includegraphics[width=7cm] {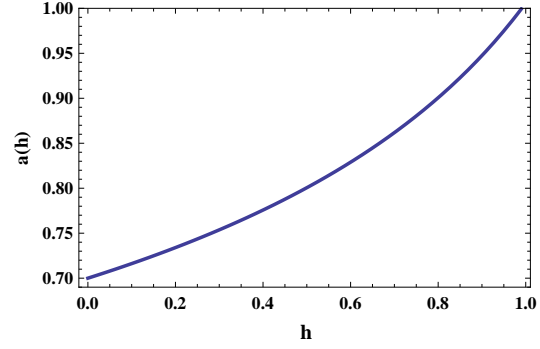}}
\caption{\label{F04} {\small   The figure displays behavior of scale factor as a function of hadron volume fraction.
We have set $\lambda = 10^{10} {\rm MeV^4}$. }}
\end{figure}

Numerically evaluated of evolution equation of the matter fraction has been plotted in Fig.(\ref{F03}). The volume fraction increases with increasing time and it reach to its maximum value, $h=1$ at about $3-9$ micro-second after the big bang, which is in agreement with the results of previous subsection. Dependence of hadron volume fraction on brane tension is clear from Fig.(\ref{F03}), which displays a faster evolution for larger value of brane tension. After that, the Universe comes to a pure hadronic phase which is discussed in next case.  \\
Evolution of scale factor of the Universe during the QHPT as a function of the hadron volume fraction is  illustrated in Fig.(\ref{F04}). It is well known that when the QHPT occurs the density of quark gluon plasma decreases but the hadron volume fraction and the scale factor of the Universe increase.

\subsubsection{Evolution of temperature in pure Hadronic Era}

At final stage of phase transition, the Universe comes to a pure hadronic phase, and cosmological fluid is described by following equation of state
\begin{equation}\label{III.28}
\rho_h=3p_h=3a_\pi T^4
\end{equation}
Using Eq.\;(\ref{II.22}), the Hubble parameter and scale factor are
\begin{equation}\label{III.29}
H = -{\dot{T} \over T} \qquad \Longrightarrow \qquad a(T) = c {T_c \over T}
\end{equation}
and from Friedmann equation (\ref{II.30}), differential equation of temperature is given by
\begin{eqnarray}\label{III.30}
\dot{T} = & -T & \biggl \{ 2\mu^4  \Big[ \chi \\
& + & \epsilon \sqrt{ \chi^2 - {1 \over 3\mu^4}
 \Big[ \mu^2 \rho_B + {\lambda \over 6}\; \rho \left( 1 + {\rho \over 2\lambda} \right) \Big] } \ \ \Bigg] \biggl \}^{\frac{1}{2}}. \nonumber
\end{eqnarray}
\begin{figure}[h]
\centerline{ \includegraphics[width=7cm] {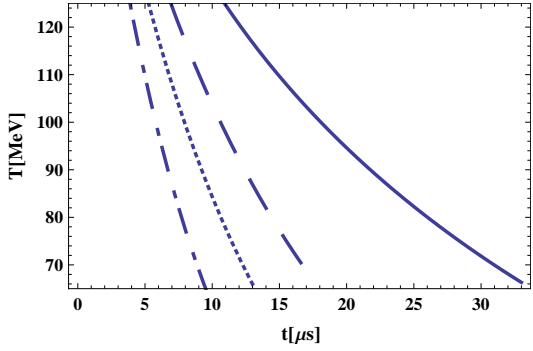}}
\caption{\label{F05} {\small   Variation of temperature $T$ versus cosmic time $t$
has been depicted for different values brane tension $\lambda$ as:
$1 \times 10^9$ (solid line), $5 \times 10^9$ (dashed line), $10 \times 10^9$
(dotted line), $20 \times 10^9$ (dotted-dashed line). The other constant parameters
are taken as: $\lambda = 10^{10} {\rm MeV^4}$, $l_{DGP} = 10^{11} {\rm MeV^{-1}}$,
$\kappa_5 = \sqrt{2\mu^2 l_{DGP}}=1 {\rm MeV^{-3/2}}$,
$\phi_0 = 2 \times 10^5 $ and $\epsilon = 1$. }}
\end{figure}
The differential equation is solved numerically and the results are depicted in Fig.(\ref{F05}). The figure indicates that the hadron phase occurs at about $10-35$ micro-second after the big bang dependence on the value of brane tension. Larger value of brane tension comes to faster temperature rate which is in agreement with the Universe expansion. The results are in good consistence with the previous results.\\

\section{Quark-Hadron Phase Transition in normal branch of the DGP model}
In previous section the problem of phase transition was investigated by using two formalism, for $\epsilon=+1$ branch of the DGP model. The case was studied in detail and the steps was explained clearly. In this  section, we are going to consider the same problem for normal branch of the DGP model, ($\epsilon=-1$ branch). The process is same as before, therefore we ignore the detail of work and go straight to results. Quark-hadron phase transition is studied by utilizing two formalisms Lattice QCD and First order phase transition respectively, and the results are mentioned as following subsections.\\

\subsection{Lattice QCD Phase Transition}
\begin{figure}[!ht]
\centerline{\includegraphics[width=7cm]{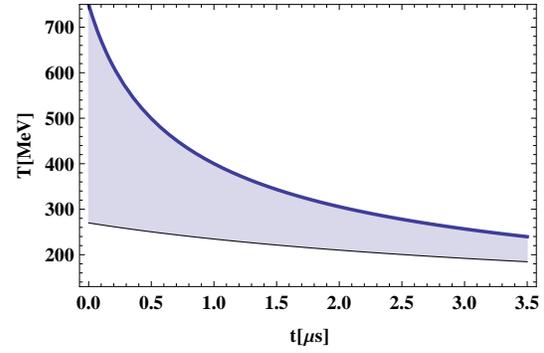}}

\caption{\label{F08a}{\small Variation of temperature $T$ versus cosmic time $t$ for high temperature regime
of the smooth crossover procedure. The constant parameters are taken as:
$\lambda = 10^{10} {\rm MeV^4}$, $l_{DGP} = 10^{11} {\rm MeV^{-1}}$,
$\kappa_5 = \sqrt{2\mu^2 l_{DGP}}=1 {\rm MeV^{-3/2}}$,
$\phi_0 = 2 \times 10^5 $ and $\epsilon = -1$.  }}

\end{figure}

\begin{figure}[!ht]
\centerline{\includegraphics[width=7cm]{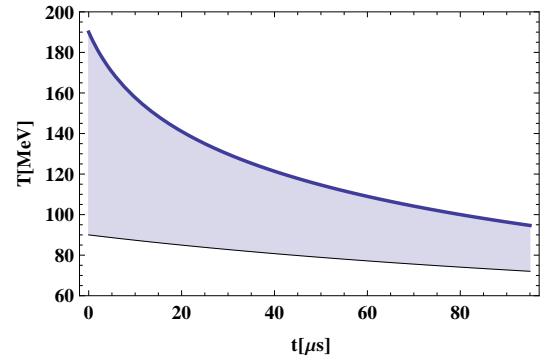}}

\caption{\label{F08b}{\small Variation of temperature $T$ versus cosmic time $t$ for low temperature regime
of the smooth crossover procedure in the DGP brane
 gravity including bulk scalar field. The constant parameters are taken as:
$\lambda = 10^{10} {\rm MeV^4}$, $l_{DGP} = 10^{11} {\rm MeV^{-1}}$,
$\kappa_5 = \sqrt{2\mu^2 l_{DGP}}=1 {\rm MeV^{-3/2}}$,
$\phi_0 = 2 \times 10^5$ and $\epsilon = -1$.  }}
\end{figure}

 Eq.(\ref{III.07}) and (\ref{III.15}) are solved numerically respectively for high and low temperature regime, in normal branch of the model. The results are plotted in Figs. \ref{F08a} and \ref{F08b}. Figure \ref{F08a} is related to high temperature regime, which shows that the temperature decreases with increasing time. The phase transition occurs at about $3-3.5$ micro-second after the big bang. The numerical results of Eq.(\ref{III.14}), related to low temperature regime, are displayed in Fig.\ref{F08b}, which expresses that phase transition occurs at about $85-90$ micro-second after the big bang. In comparison with the self-accelerating branch, ($\epsilon=+1$), { phase transition} in both high and low temperature regimes in normal branch, ($\epsilon=-1$) occurs at later times.
\subsection{First Order Phase Transition}
Quark-hadron phase transition was considered in previous section by using lattice QCD formalism. we are going to consider the problem of phase transition  in normal branch of the model by using first-order formalism. In this formalism, phase transition is studied in three step as: before, during and after phase transition. In the following, the results of three steps will be discussed.

\begin{figure}[!ht]
\centerline{\includegraphics[width=7cm]{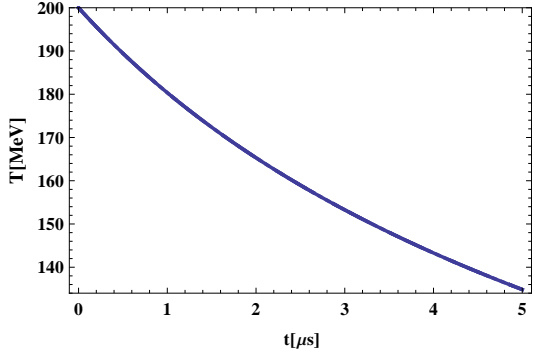}}
\caption{\label{F09a}{\small Behavior of temperature $T$ versus cosmic time $t$ for general model. The constant parameters are taken as:
$\lambda = 10^{10} {\rm MeV^4}$, $l_{DGP} = 10^{11} {\rm MeV^{-1}}$,
$\kappa_5 = \sqrt{2\mu^2 l_{DGP}}=1 {\rm MeV^{-3/2}}$,
$\phi_0 = 2 \times 10^5 $, and $\epsilon = -1$.}}
\end{figure}

\begin{figure}[!ht]
\centerline{\includegraphics[width=7cm]{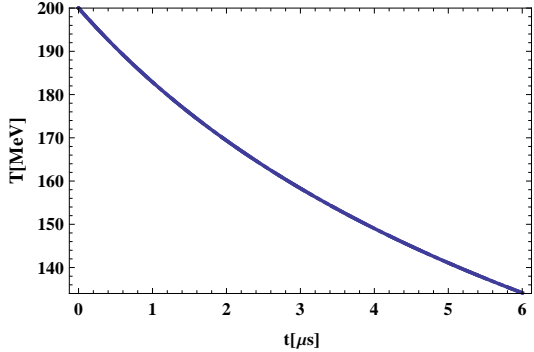}}
\caption{\label{F09b}{\small Behavior of temperature $T$ versus cosmic time $t$ for bag model. The constant parameters are taken as:
$\lambda = 10^{10} {\rm MeV^4}$, $l_{DGP} = 10^{11} {\rm MeV^{-1}}$,
$\kappa_5 = \sqrt{2\mu^2 l_{DGP}}=1 {\rm MeV^{-3/2}}$,
$\phi_0 = 2 \times 10^5$ $U(T)=B$ and $\epsilon = -1$.}}
\end{figure}

\begin{figure}[h]
\includegraphics[width=7cm]{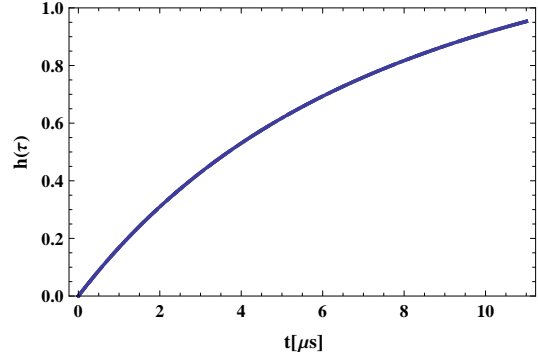}
\caption{Variation of hadron fraction parameter $h$ versus cosmic time $t$ using first order formalism. The constant parameters are taken as:
$\lambda = 10^{10} {\rm MeV^4}$, $l_{DGP} = 10^{11} {\rm MeV^{-1}}$,
$\kappa_5 = \sqrt{2\mu^2 l_{DGP}}=1 {\rm MeV^{-3/2}}$,
$\phi_0 = 2 \times 10^5$ and $\epsilon = -1$.  }\label{F10}
\end{figure}

As first step, we consider behavior of temperature before phase transition. The numerical results of Eqs.(\ref{III.21}) and (\ref{III.23}) in normal branch have been depicted {respectively} in Figs.\ref{F09a} and \ref{F09b}. It is realized that temperature is decreasing with increasing time. For general case of potential, the quark-gluon phase is finished at about $5$ micro-second after big bang, Fig.\ref{F09a}, and it finished at about $6$ micro-second after big bang for bag model Fig.\ref{F09b}. Therefore, one could realize that, according to bag model, phase of quark-gluon is finished later.
When quark-gluon is finished, the Universe comes to a phase which the energy density is a mixture of quark and hadron. Temperature and some other parameters stay constant in this era, and variation of energy densities could be described by volume fraction parameter $h$. The corresponding differential equation is explained by Eq.(\ref{III.27}), and the numerical results for $\epsilon=-1$ are plotted in Fig.\ref{F10}. Volume fraction parameter increases with increasing time, and gets to its maximum value at about $12$ micro-second after the big bang.\\

\begin{figure}[h]
\includegraphics[width=7cm]{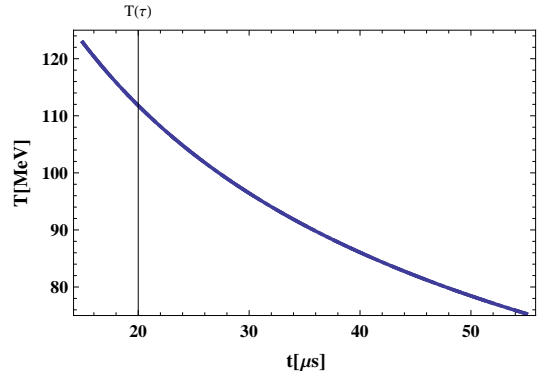}
\caption{{\small Pure hadronic temperature behavior versus cosmic time in the DGP brane
gravity including bulk scalar field by using first order formalism. The
constant parameters are taken as:
$\lambda = 10^{10} {\rm MeV^4}$, $l_{DGP} = 10^{11} {\rm MeV^{-1}}$,
$\kappa_5 = \sqrt{2\mu^2 l_{DGP}}=1 {\rm MeV^{-3/2}}$,
$\phi_0 = 2 \times 10^5$ and $\epsilon = -1$. } }\label{F11}
\end{figure}

After the volume fraction reaches the maximum value, $h=1$, the Universe enters a pure hadronic phase. Temperature decreases again, and its behavior is described by differential equation (\ref{III.30}). The differential equation is solved numerically, and the results are plotted in Fig.\ref{F11}. It shows that, hadron phase occurs at about $60$ micro-second after the big bang. \\
All of obtained results about phase transition in three steps are in good agreement with each other.
\section{Discussion and Conclusion}

The main purpose of this paper was investigation of quark-hadron phase transition problem in DGP brane-world framework including bulk scalar field, for both normal, ($\epsilon=+1$) and self-accelerating ($\epsilon=-1$) branches. Evolution of physical quantities relevant to physical description of the early times such as energy density, scale factor, and temperature, were considered during this study. After deriving {the} basic evolution equations, work has been done in two main sections, and the problem of phase transition was investigated using two popular approaches known as: 1. smooth crossover formalism and 2. first order phase transition formalism. At first, we went to the self-accelerating branch and smooth crossover approach has been selected for studying phase transition. This case { was} divided to two parts related to high and low temperature regimes. In high temperature regime, lattice QCD data for estimating trace anomaly was used to construct a realistic equation of state. As it was expected, cosmological fluid behaves like radiation in high temperature regime, and follow a simple form of equation of state. In this regime, phase transition occurs at about $2-2.5$ micro-second after the big bang. On the other hand, the trace anomaly is affected by large discretization effect in low temperature regime, therefore it is not a good model to construct an appropriate equation of state. However, Hadronic Resonance Gas (HRG) is a well-known to build a suitable equation of state in low temperature regime. The temperature differential equation was solved numerically and the results expressed a transition time at about $30-35$ micro-second after the big bang. General behavior of temperature is same for both regimes which decreases with increasing time and the Universe expansion. Phase transition depends on brane tension value, specially in low temperature regime, so that by enhancement of brane tension, phase transition occurs earlier. For normal branch, ($\epsilon=-1$), the same process was repeated. The obtained numerical results were plotted and it was realized that, phase transition in high and low temperature regimes occurs at about $3-3.5$ and $85-90$ micro-second after big bang, respectively. Therefore, there is a later transition time than self-accelerating branch.\\
The results of this case could be compare with \citep{15,25}. In \citep{15}, the authors investigated the problem in RS brane-world model including a Brans-Dicke scalar field. They found out a transition time at about few microsecond after the big bang for both regimes. In \citep{25}, the authors studied the phase transition in RS brane-world including a bulk chameleon like scalar field. This model provided a non-conservation equation of state and their results expressed a phase transition at about nano-second after the big band. However in our case, phase transition occurs at about micro-second after the big bang and displays later transition time in comparison to \citep{15} for both high and low temperature regimes.

In first order phase transition approach the temperature evolution was investigated in three steps: 1. before phase transition (quark-gluon phase); 2. during phase transition; 3. after phase transition (Hadron phase). Differential equations for temperature were solved numerically and the results depicted for both branches. The Universe effective temperature decreased with increasing time. Quark-gluon phase took place at about $2.5-4.5$ micro-second ($5$ micro-second) after the big bang for $\epsilon=+1$ ($\epsilon=-1$) branch. Phase transition for $\epsilon=+1$ ($\epsilon=-1$) branch, took place at bout $3-9$ microsecond ($12$ micro-second) after the big bang. Finally, the Universe describing by $\epsilon=+1$ ($\epsilon=-1$) branch came to a hadronic phase at about $10-35$ micro-second ($60$ micro-second) after the big bang. The acquired results in this case are in agreement with the results of previous section, and generally a consistence results were obtained during the study. Again, transition time in normal branch occurs later than self-accelerating branch.\\
In this case, our results could be compared with \citep{8}. In \citep{8} the authors consider the problem of phase transition using a RS brane-world framework. They consider the problem for different values of brane tension, and their results show a phase transition at about micro-second after the big bang. In our model, phase transition occurs at about micro-second, however later than \citep{8}. The transition time in our model was obtained for different values of brane tension and in contrast with \citep{8} we derived later transition time for larger value of brane tension.
\nocite{*}
\bibliographystyle{spr-mp-nameyear-cnd}
\bibliography{References11}
\end{document}